# Probing Magnetism in Self-Assembled Organometallic Complexes using Kondo Spectroscopy

Wantong Huang[1], Paul Greule[1], Máté Stark[1], Joris van Slageren[2], Christoph Sürgers[1], Wolfgang Wernsdorfer[1,3], Giorgio Sangiovanni[4], Christoph Wolf[5,6], Philip Willke[1,*]

[1] Physikalisches Institut, Karlsruhe Institute of Technology, 76131 Karlsruhe, Germany
[2] Institute of Physical Chemistry and Center for Integrated Quantum Science and Technology IQST, University of Stuttgart, Pfaffenwaldring 55, 70569 Stuttgart, Germany.
[3] Institute for Quantum Materials and Technologies, Karlsruhe Institute of Technology, Karlsruhe, 76021, Germany
[4] Institut für Theoretische Physik und Astrophysik and Würzburg-Dresden Cluster of Excellence ct.qmat, Universität Würzburg, Würzburg 97074, Germany
[5] Center for Quantum Nanoscience, Institute for Basic Science (IBS), Seoul, Republic of Korea.
[6] Ewha Womans University, Seoul, Republic of Korea.

* corresponding author: philip.willke@kit.edu,

## Abstract

**Control of individual spins at the atomic level holds great promise for miniaturized spintronics, quantum sensing, and quantum information processing. Both single atomic and molecular spin centers are prime candidates for these applications and are often individually addressed and manipulated using scanning tunneling microscopy (STM). In this work, we present a hybrid approach and demonstrate a robust method for self-assembly of magnetic organometallic complexes consisting of individual iron (Fe) atoms and molecules on a silver substrate using STM. We employ two types of molecules, bis(dibenzoylmethane) copper(II) [Cu(dbm)$_2$] and iron phthalocyanine (FePc). We show that in both cases the Fe atoms preferentially attach underneath the benzene ring ligand of the molecules, effectively forming an organometallic half-sandwich arene complex, Fe(C$_6$H$_6$), that is akin to the properties of metallocenes. In both situations, a molecule can be combined with up to two Fe atoms. In addition, we observe a change in the magnetic properties of the attached Fe atoms in scanning tunneling spectroscopy, revealing a distinct Kondo signature at the Fe sites. We explain the latter using density functional theory calculations, and find that the bond formation between the Fe 3$d$-orbitals and the benzene π-molecular orbitals creates a favorable situation for Kondo screening of the $d_{xz}$- and $d_{yz}$-like orbitals. Thus, this work establishes a reliable design principle for forming hybrid organometallic complexes and simultaneous tuning of their atomic spin states.**

## Introduction

In atomic-scale systems, both single atoms and molecules can be used for the miniaturization of spintronic devices[1,2] or as building blocks for quantum information processing and quantum sensors[3-5]. Both have been used in various forms in the expanding field of on-surface chemistry[6-8], which often involves scanning tunneling microscopy (STM) to image the resulting chemical structures. A prominent example, which takes advantage of both molecular and atomic structures and that has been employed here constitutes metal-organic frameworks[9-11] or hybrid complexes formed by individual atoms and molecules. While the former class of systems focusses on large interconnected assemblies, for the latter molecules and individual atoms are imaged and subsequently linked using self assembly[12,13] or atomic manipulation[14-17]. The advantage of both lies in the great variety of available building blocks. Single atoms can be combined with a molecular counterpart that can stabilize the interaction and immobilize the single atoms on a surface. Another form of prominent chemical compound, that combines metal atoms and organic molecules are formed by organometallic sandwich architectures, most notably ferrocene or nickelocene[18]. Here, they join the robust magnetic moments of $3d$ or $4f$ metal atoms with ring-shaped aromatic molecules, creating sufficient electronic hybridization between the metal atomic states via the extended π orbitals. While these are mostly synthesized in gas phase or solution[18], some examples demonstrate how they can assemble on surfaces[19,20].

If at least one of the two constituents is magnetic, a subsequent alteration of the magnetic properties is often observed, driven by e.g. exchange interaction between the spins[17], a change in the ligand field configuration[15], the spin state[14] or charge transfer[16]. Here, the change in magnetic properties is often tracked via the Kondo effect arising from the interaction between the localized spins and conduction electrons in the host metal substrate[21]. This Kondo signature results in a distinct resonance in the local density of states (LDOS) at lower temperatures and often emerges or changes when single atoms are attached to the molecular ligands[14-17]. However, atomic manipulation using the STM is often employed for their construction, which usually remains a challenging task. In this paper, we demonstrate the formation of magnetic complexes from atomic and molecular building blocks on a Ag(001) substrate utilizing self-assembly. The resulting hybrid complexes consist of one or two Fe atoms coupled to a molecule, for which we employ bis(dibenzoylmethane) copper(II) [Cu(dbm)$_2$] and iron(II) phthalocyanine (FePc). The Fe here couples to the benzene ring of the molecule and thus mimics the formation of half sandwich arene or *pianostool* complexes[18]. In all cases, we observe the emergence of a Kondo resonance located on the Fe atoms in the complex along with additional spectral features from the Fe d-orbitals. Using Density Functional Theory (DFT), we show that the bond formation between Fe 3$d$ orbitals and the π molecular orbitals of benzene follows the bond formation in conventional metallocene complexes. The resulting molecular frontier orbitals have strong $d_{xz}$ and $d_{yz}$ character and are close to half-filling, fulfilling the required conditions for a Kondo resonance of these two orbitals to emerge, which in contrast is absent in pristine Fe atoms on Ag(001).

## Results and Discussion

**Self-assembled organometallic complexes on Ag(001) surface.** Figure 1a and 1b illustrate the chemical structures of [Cu(dbm)$_2$] and FePc molecules along with corresponding STM topographic images of these molecules on Ag(001). While FePc belongs to the class of metal phthalocyanines, which have been intensively investigated on surfaces by STM techniques[22-25], [Cu(dbm)$_2$] is a promising molecular qubit candidate[26, 27], which has been rarely studied in scanning probe experiments[28, 29]. Both types of molecules have benzene rings in their outer ligand structure, which we utilize in this work as the molecular linker between individual Fe atoms and the molecule. In the STM topographies (Fig. 1a and 1b), individual [Cu(dbm)$_2$] and FePc molecules appear as butterfly-like and cross-like shapes, respectively.

Next, Fe atoms were deposited onto a cooled Ag surface. Following a gentle thermal treatment (~100 K for 10 s), we observe the formation of new hybrid complexes (Fig. 1c,d). Figure 1c displays a typical STM image of [Cu(dbm)$_2$] complexes, in which the presence of an Fe atom manifests as a local protrusion at one of the benzene rings.

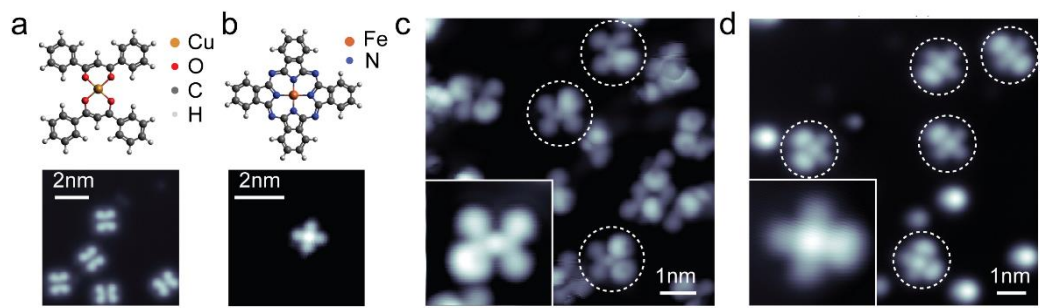

**Figure 1. Self-assembled Fe(C$_6$H$_6$) hybrid complexes on Ag(001). (a)** Structure of [Cu(dbm)$_2$] and **(b)** FePc with respective STM topographic images at the bottom showing the molecules evaporated onto a Ag(001) surface ([Cu(dbm)$_2$]: $V$ = 200mV, $I$ = 50 pA; FePc: $V$ = 100 mV, $I$ = 50 pA). **(c)** STM

topographic image showing typical complexes consisting of [Cu(dbm)$_2$] and individual Fe atoms (dashed circles; $V$ = 100 mV, $I$ = 80 pA). Here, two configurations of complexes Fe(C$_6$H$_6$)-[Cu(dbm)$_2$] (one Fe) and [Fe(C$_6$H$_6$)]$_2$-[Cu(dbm)$_2$] (two Fe) are visible. Inset: single Fe(C$_6$H$_6$)-Cu(dbm)$_2$ complex ($V$ = 200 mV, $I$ = 100 pA, image size: 2 × 2 nm$^2$) taken with a CO functionalized tip. **(d)** STM topographic image of Fe(C$_6$H$_6$)-FePc and [Fe(C$_6$H$_6$)]$_2$-FePc complexes (dashed circles) as well as several individual Fe atoms ($V$ = 100 mV, $I$ = 30 pA, image size: 10 nm). Inset: single Fe(C$_6$H$_6$)-FePc complex ($V$ = -200 mV, $I$ = 100 pA, image size: 2 × 2 nm$^2$).

The spectroscopic signature (see Fig. 2) at the protrusions along with DFT calculations (See Figure S10 in Supplementary Information), suggest that an Fe atom is located below the benzene ring arm of the molecule, denoted Fe(C$_6$H$_6$) in the following. We explored the Fe(C$_6$H$_6$) formation also for FePc, which has benzene rings in the ligand structure as well. Here, we find the same complex formation as for [Cu(dbm)$_2$] (Fig. 1d), in which one or two Fe atoms attach to the ligand of the FePc molecule. In both cases, the structure thus mimics that found in d-metal organometallic compounds, for instance metallocenes and metal–arene complexes[30]. For on-surface synthesis, a similar formation has been described for organolanthanide complexes[20] that can also show the formation of extended wires[19]. Metal–arene complexes possess strong interactions between the π-orbitals of the benzene ring and the metal d-orbitals[31, 32]. Qualitatively, this motivates the complex formation between the metal atom and the molecules here, which we continue to explore below using DFT calculations.

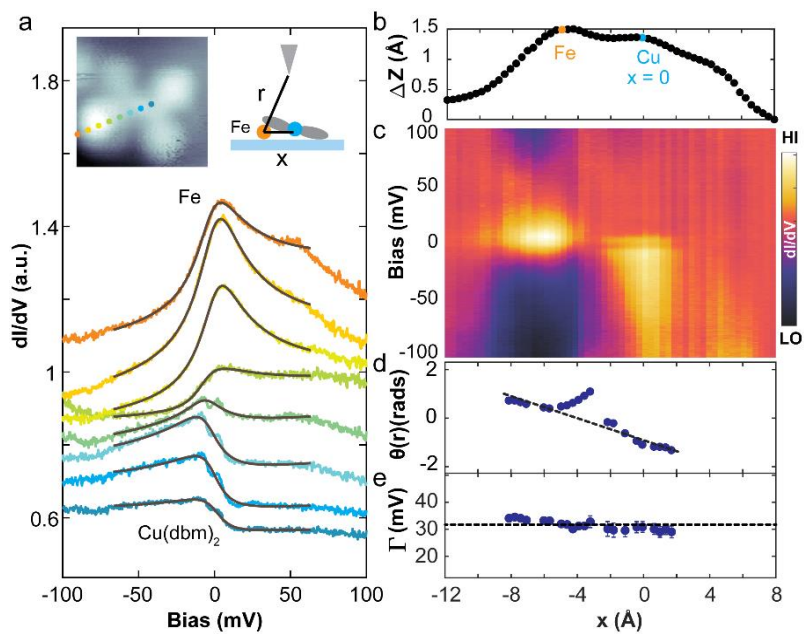

**Figure 2. Scanning Tunneling Spectroscopy on a Fe(C$_6$H$_6$)-[Cu(dbm)$_2$] complex. (a)** A series of d$I$/d$V$ spectra measured at different locations of the Fe(C$_6$H$_6$)-[Cu(dbm)$_2$] complex as indicated in the inset (Setpoint: $V$ =-100 mV, $I$ = 0.8 nA, $V_{mod}$ = 0.12 mV). The Kondo peak is fitted by a Frota function[33] (black lines). The second inset illustrates the geometric configuration between tip and complex. **(b)** Topographic profile along the line across the complex. The positions of Cu and Fe sites are labeled by blue and orange dots, respectively. **(c)** Spatially resolved d$I$/d$V$ along the same line as in (b) across the complex. **(d)** Extracted phase $\theta(r)$ and **(e)** peak width $\Gamma$ as a function of lateral position $x$ from the Frota fit [Eq. (1)]. $\theta(r)$ predominately shifts linearly from the Fe site to the Cu site (dotted line) except for a deviation observed in between. **(e)** $\Gamma$ remains mostly constant at $31.6 \pm 1.7$ mV across the complex.

**Zero bias peak on Fe(C$_6$H$_6$)-[Cu(dbm)$_2$] complex.** To investigate the arising magnetic properties of the Fe(C$_6$H$_6$) complexes, we perform position-dependent d$I$/d$V$ measurements. Figure 2a shows a series of d$I$/d$V$ spectra obtained from different locations on the Fe(C$_6$H$_6$)-[Cu(dbm)$_2$] complex. A pronounced zero bias peak (ZBP) is observed on the Fe(C$_6$H$_6$) site within the complex. The distinctive peak is completely absent in pristine [Cu(dbm)$_2$] molecules (See Figure S1 in Supplementary Information). For isolated Fe atoms, a faint zero bias feature can be found, which is however considerably weaker (See Supplementary Section 1). For magnetic systems on a conductive substrate, a ZBP is often attributed to the Kondo effect emerging from the magnetic interaction between the localized spin and the substrate bath electrons[21, 34]. To confirm the Kondo origin of the ZBP, often temperature-dependent measurements are performed[35, 36]. In the present case, the Kondo temperature $T_K$ corresponding to the observed ZBP (see below) is close to room temperature. This makes temperature-dependent measurements unfeasible, since (i) the complexes would likely become mobile and (ii) the energy resolution would not be sufficient. Here, we examine the spatial evolution of the line shape dependence at various locations within the complex, which has been employed instead of temperature dependent measurements to underpin the evidence for Kondo scattering[33]. As the tip approaches the center of [Cu(dbm)$_2$], the intensity of the Kondo peak decreases and the asymmetry changes. We fit the position-dependent Kondo peak to a Frota function[33-35]:

$$\frac{dI}{dV}(V,r) = a \cdot Im\left[-ie^{i\theta(r)}\sqrt{\frac{0.39i\Gamma}{eV-\varepsilon+0.39i\Gamma}}\right] + bV + c \quad (1)$$

where $\varepsilon$ is the position of the Kondo resonance and $\Gamma$ is the half-width at half maximum (HWHM) of the resonance, which is proportional to the Kondo temperature. Additionally, we have incorporated a linear background in Eq. (1) to account for further energy-independent tunneling processes. $\theta(r)$ represents a phase, that depends on the distance $r$ between the tip and impurity spin (see right inset of Fig. 2a) originating from the interference between the direct and indirect tunneling processes at the Kondo resonance energy[34]. The spatial evolution of the Kondo resonance along with the fitted parameters $\Gamma$ and $\theta$ in Eq. (1) is shown in Fig. 2b-e. The Kondo peak is prominently observed on the Fe site, while the line shape exhibits increasing asymmetry when moving towards the Cu site. The phase evolves predominately linearly between the Fe and Cu sites as expected for an increasing distance from a Kondo resonance[33]. We attribute the divergence observed in between the sites to either tunneling into the upper ligand, which effectively alters $r$, or to interference with higher lying spectral features discussed in Fig. 3. While also $\Gamma$ shows a slight decrease over the observed range, it stays close to 32 meV ($31.6 \pm 1.7$ meV), which suggests that the asymmetric peaks originate from the same identical impurity spin, i.e. the Fe($C_6H_6$) atom inside the complex.

**Spatially resolved d$I$/d$V$ for two types of molecules.** To further elucidate the origin of the Kondo effect, in particular the nature of the Fe orbital and spin configuration, we perform d$I$/d$V$ spectroscopy measurements at higher energies in Fig. 3. Figure 3a shows a linecut d$I$/d$V$ taken across the Fe-[Cu(dbm)$_2$] complex up to $\pm 500$mV. Within this voltage range, a spectroscopic feature at both negative and positive biases is observed (white arrows), in addition to the prominent Kondo resonance at zero bias. These high energy satellites appear close to the Fe($C_6H_6$) site within the complex. Remarkably, when we double the number of Fe atoms in the complex, both the Kondo peak and additional spectral features are doubled (Fig. 3b). In the shown configuration, the two Fe($C_6H_6$) are located on ligand sites on opposite sides of the molecule. However, we also find similar behavior in cases where two Fe atoms are located on adjacent benzene ligands (See Supplementary Section 2). We associate these additional spectral features, which are clearly related to the presence of the Fe atoms, to the lower and upper Hubbard bands (Supplementary Section 3 for further discussion and a quantitative evaluation).The latter can be understood as the removal and addition of an electron to the ground state in the presence of a local Coulomb repulsion[37]. Intriguingly, the same distinctive set of features is observed even when we change the molecule in the complex - from [Cu(dbm)$_2$] to FePc. In the single Fe($C_6H_6$)-FePc configuration (Fig. 3c) as well as the [Fe($C_6H_6$)]$_2$-FePc configuration (Fig. 3d) hosting two additional Fe atoms, we observe a Kondo peak on the Fe sites as well, accompanied again by distinctive spectral features at higher voltages. Moreover, for both [Cu(dbm)$_2$] and FePc we find additional features around zero bias in the center of the molecule. We speculate that these features correspond to the spin of the central metal atom within the molecule leading to an additional Kondo or inelastic electron tunneling spectroscopy signal (See Supplementary Section 4). These are however, difficult to differentiate[38] leaving this second spin signature evidential in this study. Regardless of the specific molecule in use, the Fe atom Kondo effect appears to emerge consistently in the presence of a benzene ring. As mentioned above, Kondo features are instead absent for individual Fe atoms on the same substrate. Moreover, a similar emergent Kondo signature has been found for hybrids of Fe atoms and NiPc[39], CuPc[39] and polyphenyl dicarbonitrile molecules[40], as well as for hybrids of Co atoms and picene[15] and Phthalocyanine[41]. Therefore, this trend suggests a general mechanism linked to the presence of the benzene ring on top of the metal atom.

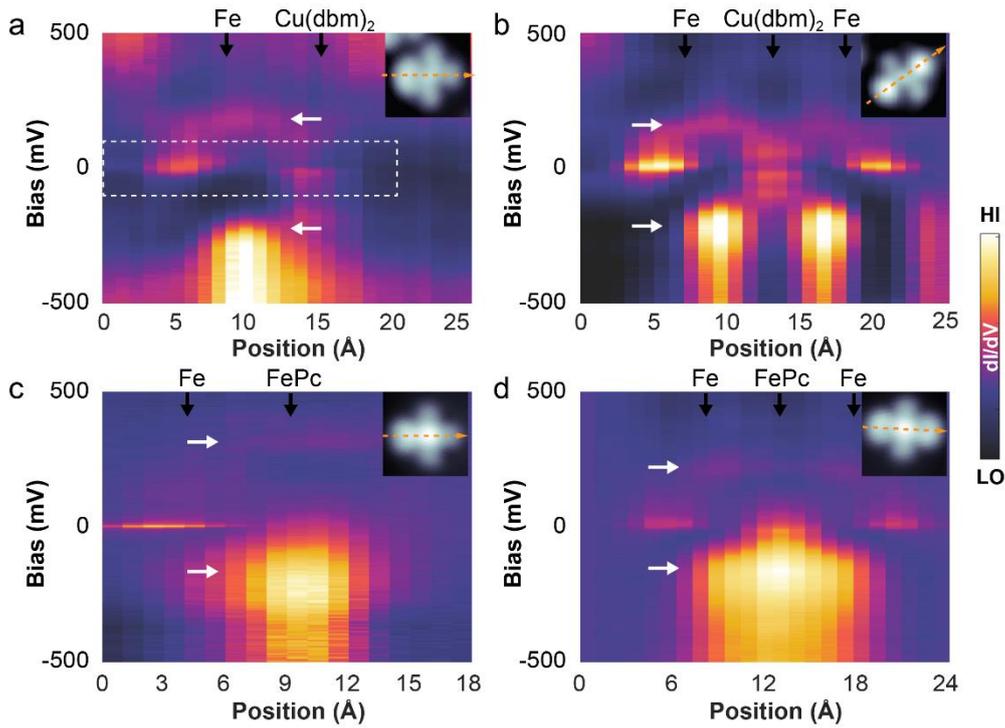

**Figure 3. Spatially resolved d$I$/d$V$ for two types of molecules. (a)** Fe(C$_6$H$_6$)-[Cu(dbm)$_2$] ($V = 500$ mV, $I = 0.5$ nA, $V_{mod} = 0.8$ mV), **(b)** [Fe(C$_6$H$_6$)]$_2$-[Cu(dbm)$_2$] ($V = 500$ mV, $I = 0.5$ nA, $V_{mod} = 0.8$ mV), **(c)** Fe(C$_6$H$_6$)-FePc ($V = 800$ mV, $I = 0.3$ nA, $V_{mod} = 0.2$ mV) and **(d)** [Fe(C$_6$H$_6$)]$_2$-FePc ($V = 500$ mV, $I = 0.15$ nA, $V_{mod} = 0.5$ mV). Spectra were taken along the orange dashed lines shown in the insets. Besides Kondo peaks around zero bias, the large bias range spectra reveal additional features at $\pm(200-300)$ mV [marked by white arrows in (a)]. The white dashed box in (a) marks the area previously shown in Fig. 2(c). Compared to (a) and (c), the number of Fe atoms is doubled in the complexes in (b) and (d). The image size for all the insets is 1.9 × 1.9 nm$^2$. For (b-c), the Ag background has been subtracted and the raw data are provided in supplementary section 3.

**DFT Calculations.** To shed light on the bonding between Fe and C$_6$H$_6$ and to explain the emergent Kondo signature on the Ag(001) surface, we employ DFT calculations. Usually, two requirements must be met for a Kondo resonance to emerge in a system composed of a paramagnetic adsorbate and a conducting substrate[21, 37]: (i) At least one of the orbitals, or a set of correlated orbitals, has to be close to being half-filled and (ii) the adsorbate orbitals should possess a sizeable hybridization with the wavefunctions of the conducting substrate[37]. The latter requirement can most easily be fulfilled for $d_{z^2}$, $d_{xz}$ and $d_{yz}$ orbitals, which can have a good overlap with the electronic wavefunctions of the underlying substrate, due to their orientation (taking the z-direction normal to the surface)[37]. However, the strength of such hybridization depends sensitively on the distance of the adsorbate to the substrate as well as the adsorption position. As a result of this hybridization, the electronic structure of the metal-adsorbate system can no longer be described in terms of discrete orbitals but involves the density of states (DOS). The contributions of adsorbate and substrate to the total DOS can be separated by using the projected DOS (PDOS) on atomic orbitals of atoms in the adsorbate and substrate. Our DFT calculations show that without the presence of a molecule individual Fe atoms are preferably adsorbed on the Ag(001) hollow site, in agreement with previous work[42]. In Figure 4a we show the resulting non-magnetic PDOS of the Fe *d*-states, as well as of the substrate Ag *s*, *p* and *d* states. The Fe *d*-states are mostly found close to the Fermi level of the combined system and are separated from the Ag *d*-states by about 2 eV. Due to the hybridization between Fe *d*-states and the substrate, the Fe *d*-states become significantly broadened. The strength of this hybridization, as a function of state energy (see Fig. 4b) can be analyzed by calculating the so-called hybridization function $\Delta(\omega)$ for a selected set of localized Wannier orbitals projected onto the states of interest, i.e. the 3*d* states of Fe[43]. The real part of this hybridization function (see Supplementary Section 5) represents the (dynamical, i.e. energy-dependent) crystal field while the imaginary part visualizes the broadening of the energy levels due to the hopping of electrons from the Fe d-orbital to the environment and back at a given frequency [44]. The hybridization for Fe 3*d* states on Ag(001) without ligands is particularly small close to the Fermi level (see inset of Fig. 4b). These values of the hybridization are incompatible with the large Kondo temperature observed in Fig. 2, indicating that the presence of the [Cu(dbm)$_2$] molecule is required for Kondo scattering to emerge. To understand how Fe in both complexes becomes Kondo-active we employ a simple model by placing a benzene ring on top of the Fe atom. Inclusion of a benzene ring does not lead to a change of the charge state of the Fe atom (see Table 1 for the occupations), i.e. no overall charge transfer is observed. However, we find that the crystal field generated by the benzene ring leads to a change of the orbital order and spin state: Orbitals with strong $d_{z2}$, $d_{xy}$ and $d_{x2-y2}$ character are now shifted and lowered in energy by ~1 eV (Fig. 4c). Moreover, the diagonal $d_{xz}$, $d_{yz}$ orbitals are now spread out over a larger energy window, which allows these orbitals to hybridize more strongly with Ag bands. This hybridization is well reflected in the strongly increased imaginary part of the hybridization function (Fig. 4d), which now shows a sizeable spectral weight around the Fermi level $E_F$ - five times larger than in the case without benzene ring. In fact, upon addition of $C_6H_6$ the states can be well described by molecular orbitals of Fe($C_6H_6$). Combining simple molecular orbital theory with our DFT (Fig. 5), we find a strong overlap between the benzene E1 orbitals and $d_{xz}$, $d_{yz}$ of the Fe, forming a strong set of $\pi$-bonds. This is also commonly the case for other metallocenes[18] and also rationalizes the stability of the Fe($C_6H_6$) complex on the surface. Similarly, the frontier orbitals are formed by an overlap between $d_{xz}$, $d_{yz}$ and the ($C_6H_6$) E1 states. Still, the frontier orbitals maintain a strong d-character (see figure caption) and are upon adsorption further hybridized with the Ag silver states (See Supplementary Section 6).

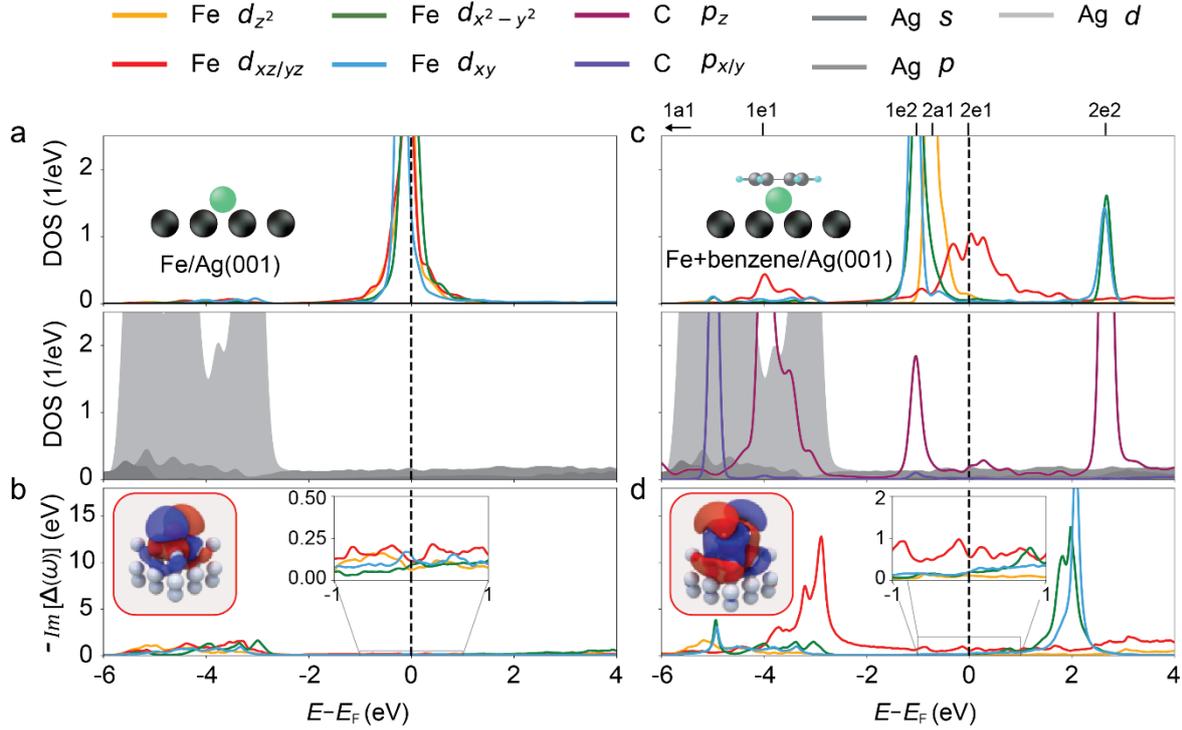

**Figure 4. DFT calculations. (a)** Projected DOS for a single Fe atom on Ag(001). Shown are the individual Fe 3$d$ orbitals as well as the substrate silver PDOS. **(b)** Imaginary part of the hybridization function, which is featureless around the Fermi level (see also inset) indicating no significant hybridization between Fe and the substrate. The second inset shows the electron density isosurface of the $d_{xz}$ orbital. **(c)** PDOS of Fe with an added benzene ring on top, Fe(C$_6$H$_6$). Here, some of the Fe 3$d$ states shift in energy, leaving the $d_{xz/yz}$-like orbitals at the Fermi level and with a sizeable hybridization (see center inset in panel d). The corresponding molecular orbitals as shown in Fig. 5 are labeled at the top. **(d)** The overall hybridization of the Fe 3$d$ states with the Ag substrate and benzene ring is substantially increased compared to the case without ring (panel b), particularly for the $d_{xz/yz}$-like orbitals (see supplementary section 7 for a discussion on the carbon atoms as source of this orbital-selective hybridization "activator"). The second inset shows the electron density isosurface of the $d_{xz}$-like orbital (2e$_1$ in Fig. 5), in which the increased overlap with the carbon atoms of the benzene ring is clearly visible.

For the Kondo effect, the role of the molecular orbitals can be further understood by performing calculations in a larger Wannier basis, which explicitly includes the carbon p-states of the benzene ring (Supplementary Section 7). In that case, we find that the $d_{z^2}$-like orbital (2a1), which often dominates Kondo scattering[37], remains largely unaffected by the presence of the benzene ring. In contrast, $d_{xz}$-, $d_{yz}$-like orbitals (2e1) show a substantial increase in hybridization near $E_F$. Therefore, the aromatic ring acts here as a selective "activator" of $d$-state hybridization. The plotted electron-density isosurfaces (inset of Fig. 4b and d) of orbitals with $d_{xz}/d_{yz}$ character (2e1) reveal that the shape of this orbital clearly changes on addition of (C$_6$H$_6$), extending further towards the benzene ring (molecular orbital formation) and the Ag substrate (Kondo effect) for Fig. 4d. To shed light on the role of the substrate, we compare Fe-benzene interaction/bonding in vacuum and on the Ag surface (Supplementary Section 6). We find that qualitatively both deliver the same results, while the surface decreases the energy splitting between *d* states and further mixes the states by hybridization with the substrate.

| Configuration | $d_{xy}$ | $d_{x2-y2}$ | $d_{xz}$ | $d_{yz}$ | $d_{z2}$ | total |
|---|---|---|---|---|---|---|
| Fe | 1.63 | 1.05 | 1.38 | 1.38 | 1.50 | 6.89 |
| Fe+benzene | 1.50 | 1.53 | 0.99 | 0.99 | 1.90 | 6.85 |

**Table 1**. Filling of the 3*d* shell for Fe on Ag(001) obtained from Lowdin charges i.e. charges obtained from projecting onto atomic basis states. The latter leads to non-integer occupations.

Due to the rearrangement in the complex, the $d_{xz}/d_{yz}$-like orbitals (2e1) are now close to being half-filled (Table 1). Thus, their broad spectral weight at $E_F$, their hybridization with the benzene and Ag substrate, and the fulfilled half-filling condition constitute strong indicators for the emergence of a Kondo resonance. Additional calculations confirm that the emergence of the Kondo resonance is not a consequence of a change in distance between the Fe and the substrate upon addition of the benzene ring. Moreover, we emphasize that reaching this conclusion via the simple modeling of a benzene ring on an Fe adatom on the Ag-substrate is a valid approximation: We confirmed by additional DFT calculations that including the full [Cu(dbm)$_2$] molecule does not change our observation (See Supplementary Section 8). Also, previous DFT calculations of Fe atoms on Cu(111) underneath polyphenyl dicarbonitrile molecules found similar results, i.e. a half-filling of $d_{xz}/d_{yz}$-like orbitals[40].

Our simple model provides therefore a generic explanation for a vast class of molecules featuring aromatic rings: Since both [Cu(dbm)$_2$] and FePc feature benzene rings, we expect in first approximation similar organometallic bond formation and subsequent hybridization effects, which explain the observation of the Kondo resonance for both molecules with configurations for 1 and 2 Fe atoms.

Notably, the Kondo peaks for Fe(C$_6$H$_6$) attached to FePc are narrower, $(4.3 \pm 0.2)$ mV, compared to those attached to [Cu(dbm)$_2$] $(31.6 \pm 1.7)$ mV. In Figure S11 (Supplementary Section 9) we show all evaluated Kondo linewidths $\Gamma$ for the different types of complexes discussed in Fig. 3. We find that in general $\Gamma$ is higher for [Cu(dbm)$_2$] than for FePc and tends to be equal or higher for molecules hosting two Fe atoms compared to hosting only one, which results in a larger spread of the observed data. This could be explained by slight changes in the geometric configuration or a small charge transfer, since both can have an effect on the shape and amplitude of the Kondo resonance[37].

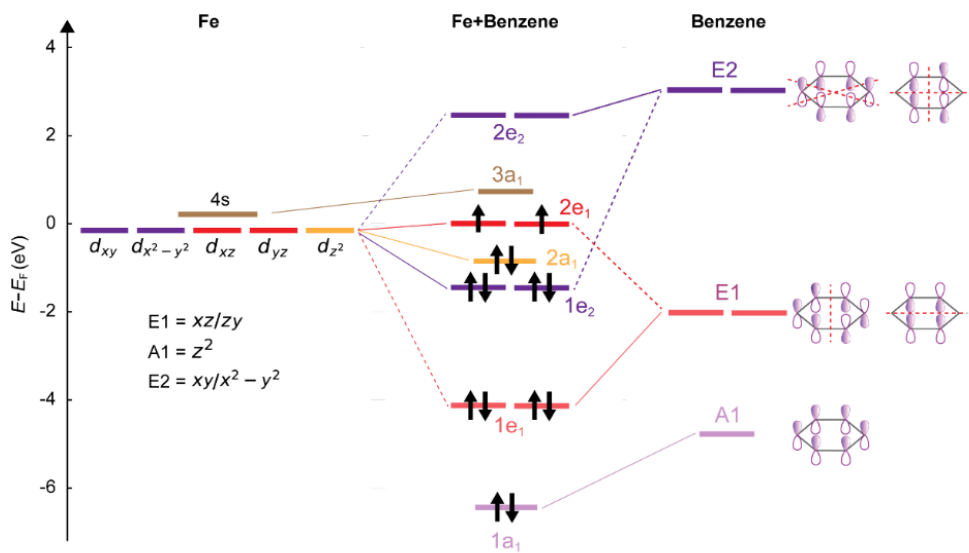

**Figure 5. Molecular orbital energy diagram of Fe and Benzene (without silver).** On the left side, Fe 3$d$ and 4$s$ states are shown with their respective symmetry group, on the right side the $\pi$ molecular orbitals of benzene. In the center the combined molecular orbitals of Fe(C$_6$H$_6$) are shown. The lowest lying states are dominated by the benzene states (1a$_1$ 90%; 1e$_1$ 84%) with some overlap of the 3$d$ and 4$s$ states. For the higher ones, the main contribution stems from the $d$ states (1e$_2$ 72%; 2a$_1$ 93%; 2e$_1$ 76%). Energies and character are obtained from DFT calculations. The addition of the Ag(001) surfaces leads to further hybridization of all states (see supplementary information 6). The relation to the PDOS is indicated in Fig. 4c.

## Conclusions

In conclusion, we demonstrated a reliable self-assembly of organometallic complexes consisting of individual Fe atoms and molecules on a silver substrate. In particular, the same kind of structure formation was observed for two different types of molecules, where in both cases Fe-arene complexes are formed by Fe atoms underneath the benzene rings of the ligands. The molecules can host up to two Fe atoms. Importantly, these hybrid complexes self-assembled on the surface by a mild thermal annealing without the use of atom manipulation. The resulting organometallic complex can be explained by overlap between $\pi$ molecular orbitals of the benzene and the Fe d-orbitals, characteristic for other metallocenes and half sandwich complexes. The subsequent change in spin structure was investigated by both STS Kondo spectroscopy and DFT calculations. The latter showed that for triggering the Kondo interaction selectively for the $d_{xz}/d_{yz}$-like orbitals the benzene ring is crucial and forms a strong overlap. We believe that the same kind of complex formation can be exploited for different kinds of aromatic ring systems and atoms other than Fe, for instance to engineer different spin states. Thus, it forms an easy and viable design principle for atomic-scale spin engineering.

## Materials and Methods

### Sample preparation

The sample preparation was carried out in-situ at a base pressure of < 6 × $10^{-10}$ mbar. The Ag(001) surface was prepared through several cycles of argon-ion sputtering and annealing through e-beam heating. [Cu(dbm)$_2$] and FePc molecules were evaporated onto the sample at room temperature using a home-built Knudsen cell, which was held at ~18W at a pressure of 1 × $10^{-9}$ mbar for 25 s and 12 s (distance ~1 cm), respectively. The deposition of Fe was carried out via electron-beam evaporation for 5 seconds onto the precooled sample transferred from the cryostat at 4 K. Subsequently, the sample was slightly thermally heated by holding it in the chamber for 20 s, resulting in the diffusion of atoms and molecules on the sample surface and the formation of organometallic complexes. All experiments were performed using an Unisoku USM1600 STM head in a closed-cycle dilution refrigerator with a base temperature of 60 mK. Before imaging, the polycrystalline Pt-Ir alloy tip was prepared by gentle poking into a clean Ag(001) surface. d*I*/d*V* spectra were acquired by the standard lock-in technique with a modulation frequency $f = 323$ Hz.

### Ab-initio Calculations

We performed periodic-cell DFT calculations using plane-wave basis and pseudopotentials in Quantum Espresso version 7.1[45, 46]. We used pseudopotentials from the PSLibrary version 1 for all atoms except Fe, where we used the PSL version 0.2.1.[47]. To construct the surface slab, we used 3 layers of Ag (a=4.16 Å) with the bottom layer frozen during relaxation. Lateral supercell sizes were chosen to ensure a separation of at least 5 Ångström between the adsorbates and their periodic images. 15 Ångström of vacuum were used to pad the cell in z-direction. All calculations used a k-grid equivalent to 63x63x63 k-points of the 1x1x1 unit cell, which is required to converge the spectral function. Spectral functions were obtained from projected Wannier orbitals using Wannier90[48] and the method described in Ref.[49-51]. All 3*d* plots in the main text were created using OVITO[52].

## Supporting Information

The Supporting Information is available free of charge at https://pubs.acs.org/doi/10.1021/acsnano.4c13172.

- dI/dV spectra of isolated Fe atoms and molecules

- Linecut across [Fe(C$_6$H$_6$)]$_2$-[Cu(dbm)$_2$]
- Characterization of the high energy satellite peaks in different complexes
- Possible spin-spin interaction in Fe(C$_6$H$_6$)-[Cu(dbm)$_2$] complexes
- Real part of the hybridization function
- Bond formation between the metal d- and benzene $\pi -$ molecular orbitals
- Role of the Benzene p states for hybridization
- Density Functional Theory Calculations of Fe(C$_6$H$_6$)-[Cu(dbm)$_2$]
- Evaluation of Kondo Temperatures

## Acknowledgments

P.W. acknowledges funding from the Emmy Noether Programme of the DFG (WI5486/1-1) and financing from the Baden Württemberg Foundation Program on Quantum Technologies (Project AModiQuS). C.W. acknowledges support from the Institute for Basic Science under Grant IBS-R27-D1. P.G. and P.W. acknowledges financial support from the Hector Fellow Academy (Grant No. 700001123). G.S. was supported by the Deutsche Forschungsgemeinschaft (DFG, German Research Foundation) through the Würzburg-Dresden Cluster of Excellence on Complexity and Topology in Quantum Matter "ct.qmat" (Project No. 390858490, EXC 2147) and through the DFG research unit "QUAST" FOR 5249-449872909 (Project P5). We thank Domenico Di Sante for useful discussions.

## References

(1) Aradhya, S. V.; Venkataraman, L. Single-molecule junctions beyond electronic transport. *Nature Nanotechnology* **2013**, *8* (6), 399-410.
(2) Choi, D.-J.; Lorente, N.; Wiebe, J.; von Bergmann, K.; Otte, A. F.; Heinrich, A. J. Colloquium: Atomic spin chains on surfaces. *Reviews of Modern Physics* **2019**, *91* (4), 041001.
(3) Chen, Y.; Bae, Y.; Heinrich, A. J. Harnessing the Quantum Behavior of Spins on Surfaces. *Advanced Materials* **2023**, *35* (27), 2107534.
(4) Gaita-Ariño, A.; Luis, F.; Hill, S.; Coronado, E. Molecular spins for quantum computation. *Nature Chemistry* **2019**, *11* (4), 301-309.
(5) Moreno-Pineda, E.; Wernsdorfer, W. Measuring molecular magnets for quantum technologies. *Nature Reviews Physics* **2021**, *3* (9), 645-659.
(6) Clair, S.; de Oteyza, D. G. Controlling a Chemical Coupling Reaction on a Surface: Tools and Strategies for On-Surface Synthesis. *Chemical Reviews* **2019**, *119* (7), 4717-4776.
(7) Shen, Q.; Gao, H.-Y.; Fuchs, H. Frontiers of on-surface synthesis: From principles to applications. *Nano Today* **2017**, *13*, 77-96.
(8) Sun, Q.; Zhang, R.; Qiu, J.; Liu, R.; Xu, W. On-Surface Synthesis of Carbon Nanostructures. *Advanced Materials* **2018**, *30* (17), 1705630.
(9) Dong, L.; Gao, Z. A.; Lin, N. Self-assembly of metal–organic coordination structures on surfaces. *Progress in Surface Science* **2016**, *91* (3), 101-135.
(10) Lin, T.; Kuang, G.; Wang, W.; Lin, N. Two-Dimensional Lattice of Out-of-Plane Dinuclear Iron Centers Exhibiting Kondo Resonance. *ACS Nano* **2014**, *8* (8), 8310-8316.
(11) Iancu, V.; Braun, K.-F.; Schouteden, K.; Van Haesendonck, C. Inducing Magnetism in Pure Organic Molecules by Single Magnetic Atom Doping. *Physical Review Letters* **2014**, *113* (10), 106102.

(12) Röckert, M.; Franke, M.; Tariq, Q.; Steinrück, H.-P.; Lytken, O. Evidence for a precursor adcomplex during the metalation of 2HTPP with iron on Ag(100). *Chemical Physics Letters* **2015**, *635*, 60-62.
(13) Smykalla, L.; Shukrynau, P.; Zahn, D. R. T.; Hietschold, M. Self-Metalation of Phthalocyanine Molecules with Silver Surface Atoms by Adsorption on Ag(110). *The Journal of Physical Chemistry C* **2015**, *119* (30), 17228-17234.
(14) Wahl, P.; Diekhöner, L.; Wittich, G.; Vitali, L.; Schneider, M. A.; Kern, K. Kondo Effect of Molecular Complexes at Surfaces: Ligand Control of the Local Spin Coupling. *Physical Review Letters* **2005**, *95* (16), 166601.
(15) Zhou, C.; Shan, H.; Li, B.; Zhao, A.; Wang, B. Engineering hybrid Co-picene structures with variable spin coupling. *Applied Physics Letters* **2016**, *108* (17).
(16) Choi, T.; Badal, M.; Loth, S.; Yoo, J. W.; Lutz, C. P.; Heinrich, A. J.; Epstein, A. J.; Stroud, D. G.; Gupta, J. A. Magnetism in Single Metalloorganic Complexes Formed by Atom Manipulation. *Nano Letters* **2014**, *14* (3), 1196-1201.
(17) Wegner, D.; Yamachika, R.; Zhang, X.; Wang, Y.; Baruah, T.; Pederson, M. R.; Bartlett, B. M.; Long, J. R.; Crommie, M. F. Tuning Molecule-Mediated Spin Coupling in Bottom-Up-Fabricated Vanadium-Tetracyanoethylene Nanostructures. *Physical Review Letters* **2009**, *103* (8), 087205.
(18) Crabtree, R. H. *The Organometallic Chemistry of the Transition Metals*; John Wiley & Sons, 2009.
(19) Huttmann, F.; Schleheck, N.; Atodiresei, N.; Michely, T. On-Surface Synthesis of Sandwich Molecular Nanowires on Graphene. *Journal of the American Chemical Society* **2017**, *139* (29), 9895-9900.
(20) Mathialagan, S. K.; Parreiras, S. O.; Tenorio, M.; Černa, L.; Moreno, D.; Muñiz-Cano, B.; Navío, C.; Valvidares, M.; Valbuena, M. A.; Urgel, J. I.; et al. On-Surface Synthesis of Organolanthanide Sandwich Complexes. *Advanced Science* **2024**, *n/a* (n/a), 2308125.
(21) Ternes, M.; Heinrich, A. J.; Schneider, W.-D. Spectroscopic manifestations of the Kondo effect on single adatoms. *Journal of Physics: Condensed Matter* **2009**, *21* (5), 053001.
(22) Mugarza, A.; Robles, R.; Krull, C.; Korytár, R.; Lorente, N.; Gambardella, P. Electronic and magnetic properties of molecule-metal interfaces: Transition-metal phthalocyanines adsorbed on Ag(100). *Physical Review B* **2012**, *85* (15), 155437.
(23) Tsukahara, N.; Noto, K.-i.; Ohara, M.; Shiraki, S.; Takagi, N.; Takata, Y.; Miyawaki, J.; Taguchi, M.; Chainani, A.; Shin, S.; et al. Adsorption-Induced Switching of Magnetic Anisotropy in a Single Iron(II) Phthalocyanine Molecule on an Oxidized Cu(110) Surface. *Physical Review Letters* **2009**, *102* (16), 167203.
(24) Gao, L.; Ji, W.; Hu, Y. B.; Cheng, Z. H.; Deng, Z. T.; Liu, Q.; Jiang, N.; Lin, X.; Guo, W.; Du, S. X.; et al. Site-Specific Kondo Effect at Ambient Temperatures in Iron-Based Molecules. *Physical Review Letters* **2007**, *99* (10), 106402.
(25) Minamitani, E.; Tsukahara, N.; Matsunaka, D.; Kim, Y.; Takagi, N.; Kawai, M. Symmetry-Driven Novel Kondo Effect in a Molecule. *Physical Review Letters* **2012**, *109* (8), 086602.
(26) Ciccullo, F.; Glaser, M.; Sättele, M. S.; Lenz, S.; Neugebauer, P.; Rechkemmer, Y.; van Slageren, J.; Casu, M. B. Thin film properties and stability of a potential molecular quantum bit based on copper(ii). *Journal of Materials Chemistry C* **2018**, *6* (30), 8028-8034, 10.1039/C8TC02610F.
(27) Lenz, S.; Bader, K.; Bamberger, H.; van Slageren, J. Quantitative prediction of nuclear-spin-diffusion-limited coherence times of molecular quantum bits based on copper(ii). *Chemical Communications* **2017**, *53* (32), 4477-4480, 10.1039/C6CC07813C.
(28) Leoni, T.; Guillermet, O.; Walch, H.; Langlais, V.; Scheuermann, A.; Bonvoisin, J.; Gauthier, S. Controlling the Charge State of a Single Redox Molecular Switch. *Physical Review Letters* **2011**, *106* (21), 216103.
(29) Walch, H.; Leoni, T.; Guillermet, O.; Langlais, V.; Scheuermann, A.; Bonvoisin, J.; Gauthier, S. Electromechanical switching behavior of individual molecular complexes of Cu and Ni on NaCl-covered Cu(111) and Ag(111). *Physical Review B* **2012**, *86* (7), 075423.
(30) Atkins, P. *Shriver and Atkins' inorganic chemistry*; Oxford University Press, USA, 2010.

(31) Pampaloni, G. Aromatic hydrocarbons as ligands. Recent advances in the synthesis, the reactivity and the applications of bis(η6-arene) complexes. *Coordination Chemistry Reviews* **2010**, *254* (5), 402-419.
(32) Duncan, M. A. Structures, energetics and spectroscopy of gas phase transition metal ion–benzene complexes. *International Journal of Mass Spectrometry* **2008**, *272* (2), 99-118.
(33) Prüser, H.; Wenderoth, M.; Dargel, P. E.; Weismann, A.; Peters, R.; Pruschke, T.; Ulbrich, R. G. Long-range Kondo signature of a single magnetic impurity. *Nature Physics* **2011**, *7* (3), 203-206.
(34) Frank, S.; Jacob, D. Orbital signatures of Fano-Kondo line shapes in STM adatom spectroscopy. *Physical Review B* **2015**, *92* (23), 235127.
(35) Mishra, S.; Beyer, D.; Eimre, K.; Kezilebieke, S.; Berger, R.; Gröning, O.; Pignedoli, C. A.; Müllen, K.; Liljeroth, P.; Ruffieux, P.; et al. Topological frustration induces unconventional magnetism in a nanographene. *Nature Nanotechnology* **2020**, *15* (1), 22-28.
(36) Jacob, D. Temperature evolution of the Kondo peak beyond Fermi liquid theory. *Physical Review B* **2023**, *108* (16), L161109.
(37) Kügel, J.; Karolak, M.; Senkpiel, J.; Hsu, P.-J.; Sangiovanni, G.; Bode, M. Relevance of hybridization and filling of 3d orbitals for the Kondo effect in transition metal phthalocyanines. *Nano letters* **2014**, *14* (7), 3895-3902.
(38) Noei, N.; Mozara, R.; Montero, A. M.; Brinker, S.; Ide, N.; Guimarães, F. S. M.; Lichtenstein, A. I.; Berndt, R.; Lounis, S.; Weismann, A. Manipulating the Spin Orientation of Co Atoms Using Monatomic Cu Chains. *Nano Letters* **2023**, *23* (19), 8988-8994.
(39) Krull, C. *Electronic structure of metal phthalocyanines on Ag (100)*; Springer Science & Business Media, 2013.
(40) Pacchioni, G. E.; Pivetta, M.; Gragnaniello, L.; Donati, F.; Autès, G.; Yazyev, O. V.; Rusponi, S.; Brune, H. Two-Orbital Kondo Screening in a Self-Assembled Metal–Organic Complex. *ACS Nano* **2017**, *11* (3), 2675-2681.
(41) Zhu, L.; Li, B.; Dong, L.; Feng, W.; Zhao, A.-d.; Wang, B. Controlling metalation reaction of phthalocyanine with cobalt at single-molecule level on Au(111) surface. *Chinese Journal of Chemical Physics* **2021**, *34* (4), 419-428.
(42) Gardonio, S.; Karolak, M.; Wehling, T. O.; Petaccia, L.; Lizzit, S.; Goldoni, A.; Lichtenstein, A. I.; Carbone, C. Excitation Spectra of Transition-Metal Atoms on the Ag (100) Surface Controlled by Hund's Exchange. *Physical Review Letters* **2013**, *110* (18), 186404.
(43) Gull, E.; Millis, A. J.; Lichtenstein, A. I.; Rubtsov, A. N.; Troyer, M.; Werner, P. Continuous-time Monte Carlo methods for quantum impurity models. *Reviews of Modern Physics* **2011**, *83* (2), 349-404.
(44) Bahlke, M. P.; Schneeberger, M.; Herrmann, C. Local decomposition of hybridization functions: Chemical insight into correlated molecular adsorbates. *The Journal of Chemical Physics* **2021**, *154* (14).
(45) Giannozzi, P.; Baroni, S.; Bonini, N.; Calandra, M.; Car, R.; Cavazzoni, C.; Ceresoli, D.; Chiarotti, G. L.; Cococcioni, M.; Dabo, I.; et al. QUANTUM ESPRESSO: a modular and open-source software project for quantum simulations of materials. *Journal of Physics: Condensed Matter* **2009**, *21* (39), 395502.
(46) Giannozzi, P.; Andreussi, O.; Brumme, T.; Bunau, O.; Buongiorno Nardelli, M.; Calandra, M.; Car, R.; Cavazzoni, C.; Ceresoli, D.; Cococcioni, M.; et al. Advanced capabilities for materials modelling with Quantum ESPRESSO. *Journal of Physics: Condensed Matter* **2017**, *29* (46), 465901.
(47) Dal Corso, A. Pseudopotentials periodic table: From H to Pu. *Computational Materials Science* **2014**, *95*, 337-350.
(48) Mostofi, A. A.; Yates, J. R.; Pizzi, G.; Lee, Y.-S.; Souza, I.; Vanderbilt, D.; Marzari, N. An updated version of wannier90: A tool for obtaining maximally-localised Wannier functions. *Computer Physics Communications* **2014**, *185* (8), 2309-2310.
(49) Amadon, B.; Lechermann, F.; Georges, A.; Jollet, F.; Wehling, T. O.; Lichtenstein, A. I. Plane-wave based electronic structure calculations for correlated materials using dynamical mean-field theory and projected local orbitals. *Physical Review B* **2008**, *77* (20), 205112.
(50) Karolak, M. Electronic correlation effects in transition metalsystems: From bulk crystals to nanostructures. Staats-und Universitätsbibliothek Hamburg Carl von Ossietzky, 2013.

(51) Odobesko, A.; Di Sante, D.; Kowalski, A.; Wilfert, S.; Friedrich, F.; Thomale, R.; Sangiovanni, G.; Bode, M. Observation of tunable single-atom Yu-Shiba-Rusinov states. *Physical Review B* **2020**, *102* (17), 174504.
(52) Stukowski, A. Visualization and analysis of atomistic simulation data with OVITO–the Open Visualization Tool. *Modelling and Simulation in Materials Science and Engineering* **2010**, *18* (1), 015012.